\documentstyle[prb,multicol,aps,epsfig]{revtex}
\begin{document}
\title{Single parameter scaling in 1-{\em D} Anderson localization. Exact
analytical solution}
\author{Lev I. Deych}
\address{Physics Department, Seton Hall University, South Orange, NJ 07052}
\author{A. A. Lisyansky}
\address{Physics Department, Queens College of CUNY, Flushing, NY 11367}
\author{B. L. Altshuler}
\address{Physics Department, Princeton University and NEC Research Institute,\\
Princeton NJ 08540}
\maketitle

\begin{abstract}
The variance of the Lyapunov exponent is calculated exactly in the
one-dimensional Anderson model with random site energies distributed
according to the Cauchy distribution. We derive an exact analytical
criterion for the validity of the single parameter scaling  in this
model. According to this criterion, states with energies from the
conduction band of the underlying non-random system satisfy single parameter scaling when
disorder is small enough. At the same time, single parameter scaling is not valid for states
close to band boundaries and those outside of the original spectrum even in
the case of small disorder. The obtained results are applied to the
Kronig-Penney model with the potential in the form of periodically positioned 
$\delta$-functions with random strengths. We show that an increase in the
disorder can restore single parameter scalingbehavior for states from band-gaps of this
model.
\end{abstract}


\begin{multicols}{2}

\section{Introduction}

The hypothesis of single-parameter scaling (SPS) in the context of
transport properties of disordered conductors was introduced in Ref. %
\onlinecite{gang}. It was suggested that scaling properties of the conductance $%
g $ are determined by a single parameter, the conductance itself, through a
scaling equation 
\begin{equation}
\frac{d(\ln g)}{d(\ln L)}=\beta (g),  \label{scaling}
\end{equation}
where $L$ is the size of a sample. The nature of the scaling conductance $g$
was debated for some time until it was understood\cite{Anderson} that
scaling in the theory of localization must be interpreted in terms of the
entire distribution function of conductivity rather then in terms of its
momentums. SPS in this case means that the distribution function of $g$ is
fully determined by a single parameter, which obeys scaling equation (\ref
{scaling}). In Ref. \onlinecite{Anderson}, which was concerned with scaling
properties of one-dimensional disordered conductors, Anderson, et al.
proposed the parameter 
\begin{equation}
\tilde{\gamma}(L)=\frac{1}{2L}$ $\ln \left( 1+\displaystyle{\frac{1}{g}}\right) \label{gamma1}
\end{equation}
as a scaling parameter suitable to describe the fluctuations of the
conductivity. In the limit $L\rightarrow \infty $ the introduced parameter
takes a non-random value, $\gamma $, which is the inverse localization
length, $l_{loc}$, or the Lyapunov exponent (LE) characterizing the spatial distribution
of electron's wave functions.\cite{Oseledets,LGP} It was suggested in Ref. %
\onlinecite{Anderson} that the introduced parameter has a normal
distribution and does not exhibit anomalously large fluctuations.
Calculations carried out in Ref. \onlinecite{Anderson} showed that the
variance of LE, $\sigma ^{2}$, scales according to the law of large numbers $%
\sigma ^{2}\sim 1/L$, and is related in a universal way to LE: \cite{note1}
\begin{equation}
\sigma ^{2}=\gamma/L.  \label{SPS}
\end{equation}
This relation is the essence of SPS in the case of strong localization, as it
presumes that two parameters of the normal distribution of LE are reduced to
one scaling parameter, $\gamma $.  According to the scaling theory (see
Ref. \onlinecite{gang} and references therein) almost all states in
one-dimensional systems are localized, and these systems are, therefore,
always in the regime of strong localization in the asymptotic limit $L\gg
l_{loc}$. 

Eq. (\ref{SPS}) was first derived in Ref. \onlinecite{Anderson} within  the approximation known as the random phase hypothesis,
which assumes  that there exists a microscopic length scale over which phases of complex transmission and reflection coefficients become completely randomized.  Under similar assumptions, Eq. (\ref{SPS})
was rederived later by several authors for a number of different models.\cite
{Abrikosov,Melnikov,Pendry,Pichard1,Heinrichs,Mielke}

Landauer's representation of the conductance in terms of transmission
coefficients for different scattering channels\cite{Landauer,Fisher} reduces the
study of the conductance in quasi-one-dimensional wires to the analysis of
scattering or transfer matrices. Within the transfer-matrix approach \cite
{transfer-matrix} the problem is further reduced to the study of statistical
properties of the products of random matrices (see Ref. \onlinecite
{Beenakker} and references therein). In this context, the
self-averaging of LE and its normal distribution in the asymptotic limit $%
L\rightarrow \infty $ are rigorously established mathematical facts. \cite
{Oseledets,Furstenberg,Mello} SPS expressed by Eq. (%
\ref{SPS}) was also established in the limit of strong localization for a
quasi-one-dimensional geometry in Ref.\onlinecite{Dorokhov,Pichard} with the use
of the Dorokhov-Mello-Pereyra-Kumar equation.\cite{DPMK}

The SPS hypothesis has also been verified in the regime of weak localization,
which exists in the conducting phase of three dimensional conductors and in
the limit of a large number of scattering channels in the
quasi-one-dimensional geometry. In contrast with the case of strong localization, the distribution of the conductance (rather than the logarithm of the conductance) 
has the Gaussian form with two independent parameters. The variance of the
conductance, however, was found to be a universal number\cite{UCF1,UCF2}
leaving one again with a single scaling parameter - the average conductance. The
universal conductance fluctuations, which were first discovered in
three-dimensional conductors within a diagrammatic approach, \cite{UCF1,UCF2}
were later reinterpreted from the point of view of the random matrix theory in
a quasi-one-dimensional geometry (see for review Ref. \onlinecite
{Beenakker}).

The scaling properties of conductivity have also been studied
numerically by a number of authors. The general concepts of the scaling theory expressed by Eq. (\ref
{scaling}) were verified by means of Green's function and transfer-matrix
approaches generalized for two- and three-dimensional systems in Ref.\onlinecite
{Cramer}. The log-normal distribution of the conductance and SPS Eq. (\ref
{SPS}) has been confirmed for the Anderson model (AM) in numerical simulations in
Ref.  \onlinecite{Pichard}. In the one-dimensional situation, the existence of SPS
has been also obtained in simulations of AM with correlated
disorder, \cite{Oliveira} and scalar wave propagation in superlattices with
different models of randomness.\cite{Tamura,Dima1}

The zero energy state in one-dimensional models with off-diagonal disorder
(random hopping models) represents a special case. These models demonstrate a
delocalization transition in the vicinity of zero energy \cite{off-diagonal}
contrary to the conclusion of the scaling theory that such a
transition is absent in one-dimensional systems. The SPS relation (\ref{SPS}) between
the standard deviation of LE and its mean value is also violated in this case.\cite
{off-diagonal1} Unusual properties of this model are due to a so called
chiral symmetry, which is characteristic of the state with $E=0$ in this
model. In further discussion we will ignore this special case and refer
to regular situations, which include models with diagonal disorder and
random hopping models away from the critical $E=0$ point.

Simultaneously with numerous confirmations of the existence of SPS, the
limits of its validity have been the subject of intensive discussions (see, for
example, Ref.\onlinecite{Shapiro} and references therein). As we mentioned above, the original condition
for SPS, postulated in Ref. \onlinecite{Anderson}, invokes the hypothesis of the
phase randomization. This hypothesis implies that the phases of complex
transmission and reflection coefficients become completely randomized at the
distances much smaller than  the localization length. Phase randomization
was numerically studied for AM in Ref. \onlinecite{Stone}, where it was
shown that for small disorder the phases indeed become uniformly distributed
at a scale much shorter than the localization length. This does not happen,
however, for states in the center of the original conduction band. Numerical
calculation of AM in Ref.\onlinecite{Stone} and analytical
consideration of a model with periodically positioned scatterers in Ref. \onlinecite
{Dmitriev} obtained a nonuniform phase distribution for such states. It was
shown in Ref. \onlinecite{Dmitriev} that for the states at the center of the band there exists a new length, a phase relaxation
length, $l_{\varphi}$. As soon as the length of the sample exceeds $l_{\varphi}$, the phase distribution  
approaches a stationary but nonuniform form. Under certain conditions, the
relaxation length rather than the mean-free-path was found to determine the
localization length. At the same time, neither analytical nor numerical studies of the states at the center of the conduction band found violations of SPS. These results cast
doubt upon the relevance of phase randomization for SPS. 
In this paper we show that the condition for SPS is not phase randomization.

When local disorder is strong, the phase distribution was found to never become
uniform,\cite{Stone} and the probability distribution of LE, in this
case, is controlled by two independent parameters.\cite{Shapiro,Stone} It was indicated, 
\cite{Stone} however, that even in the case of an extremely nonuniform phase
distribution, the deviations from SPS are rather limited.

The hypothesis of phase randomization lies at the foundation of all existing
theoretical approaches to statistical properties of conductance, including
those based upon random matrix theory.\cite{Beenakker} An additional
requirement crucial for Eq. (\ref{SPS}) can be called ``local weakness of disorder." In
calculations based upon the random matrix theory, \cite{Mello,DPMK} this
requirement is set as a limit when the
cross-section of each individual scatterer tends to zero, while the density of
the scatterers tends to infinity keeping the localization length constant.
It is commonly believed that in the regime of strong localization ($l_{loc} \ll L$), SPS holds provided that the local disorder is weak, so that the localization length exceeds all microscopic length scales of the model.  Increase of the disorder leads to reduction of the localization length, and eventually violates SPS.

Results which apparently contradict this well established understanding 
of the crossover between SPS and statistics with two independent
parameters, were recently reported in Ref. \onlinecite{Dima2}. The system
considered in Ref. \onlinecite{Dima2} belongs to the class of
Kronig-Penney-like models (KPM), which have been intensively studied 
(see, for example, Refs. \onlinecite{LGP,Erdos,Altshuler} and references
therein). The original spectrum of KPM contains multiple bands separated by
band-gaps. Disorder not only localizes states within the original pass bands
but also creates tails of localized states in former band gaps.\cite{LGP} According to
Ref. \onlinecite{Dima2} the spectrum of the system is divided into two
groups of states with different scaling behavior: SPS holds for states from the conduction bands
of the initial spectrum, and is violated for states from
initial band-gaps. Moreover, this violation of SPS for band-gap states
occurs even for weak disorder, and turns out to be much more dramatic than the phase randomization approach would predict.\cite{Stone}

Occurrence of states outside of the initial conduction bands is known to be a model independent phenomenon. It seems plausible, therefore, that the coexistence of SPS and non-SPS states found in Ref.%
\onlinecite{Dima2} is not a particular feature of KPM but rather a general property of quantum disordered systems.

The main objective of the present paper is to re-examine the problem of
scaling properties of conductance in one-dimensional systems and to derive
SPS Eq. (\ref{SPS}) without the assumption of phase randomization. This
calculation allows us to formulate a ``correct'' criterion for SPS, and to understand the
nature of its violation reported in Ref.\onlinecite{Dima2}. The main results of this paper were outlined in Ref. \onlinecite{PRL}.  

The paper is organized as follows: In Sec. II we formulate the model within which we calculate the variance of LE.  The details of the calculations are presented in Sec. III.  The new criterion of SPS is derived and analyzed in Sec. IV.  In that section we also complement our analytical
calculations with numerical simulations of a more generic model.  Comparison
with the latter helps us to distinguish between universal features of
our results and those specific to the selected model.  The transition between SPS and non-SPS states is discussed in Sec. V.  We conclude in Sec. VI.

\section{The description of the model}

Let us consider a one-dimensional tight binding model with
diagonal disorder, which is described by the following equations of motion

\begin{equation}
\psi _{n+1}+\psi _{n-1}-U_{n}\psi _{n}=0,  \label{model}
\end{equation}
where $\psi _{n}$ represents the wave function of the system at the $n$-th site.
In Eq. (\ref{model}) 
the hopping integral is chosen to be equal
to one, so it sets the energy scale in the system. The concrete meaning of $%
U_{n}$ depends upon the interpretation of the model (\ref{model}). There are
two apparently different models that can be described by Eq. (\ref{model}). In the first, this equation represent a classical
AM, with $U_{n}$ defined as 
\begin{equation}
U_{n}=-E+\epsilon _{n},  \label{UAndreson}
\end{equation}
where $E$ is the energy of a particle, and $\epsilon _{n}$ is the random site
energy. Second, it can be shown (see, for example, \cite{LGP}) that the
Schr\"{o}dinger equation for KPM with a random potential
formed by periodically positioned $\delta $-functions with random strengths, 
$V_{n}$, also reduces to the form (\ref{model}) with $\psi _{n}$
being the values of the eigenfunctions at the sites occupied by the $\delta $%
-potentials. In this case, $U_{n}$ is defined as follows:\cite{LGP}
\begin{equation}
U_{n}=2\cos (ka)+\frac{V_{n}}{k}\sin (ka),  \label{UKronig}
\end{equation}
where $k=\sqrt{E}$ is the energy variable and $a$ is the period of the
structure. To be able to obtain an exact analytical solution, we assume that parameters $\epsilon _{n}$ or $V_{n}$ are
distributed with the Cauchy probability density (the Lloyd model):
\begin{equation}
P_{C}(x)=\frac{1}{\pi }\frac{\Gamma }{\Gamma ^{2}+(x-x_{0})^{2}},  \label{P}
\end{equation}
where $x_{0}=0$ or $V_{0}$ for AM or KPM,
respectively. Parameters $x_{0}$ and $\Gamma $ represent the mean
value of the random variable $x$ and the width of the distribution,
respectively. Although $\Gamma $ characterizes the strength of disorder in the
system, it cannot be interpreted as a second moment of the distribution (%
\ref{P}), because the latter does not exist. AM with the probability
distribution (\ref{P}) is one of the first models where LE was evaluated
exactly.\cite{LGP,Ishii} The probability distribution of parameters $U_{n}$,
which enter equations of motion (\ref{model}), has the same form as Eq. (\ref{P}) with the following parameters:
\begin{equation}
\left\langle U_{n}\right\rangle =U_{0}=\left\{ 
\matrix{
2\cos (ka)+{\displaystyle{V_{0} \over k}}
\sin (ka), & \text{KPM} \cr
E,&\text{ AM}}
\right. \text{ }  \label{UAV}
\end{equation}

\bigskip

\begin{equation}
\Gamma _{U}=\left\{ 
\matrix{
{\displaystyle{\Gamma  \over k}}\left| \sin (ka)\right|, &\text{ KPM} \cr 
\Gamma. &\text{ AM}}
\right.  \label{width}
\end{equation}

In the absence of disorder, the energy spectrum of the model is
determined by the condition: $\left| U_{0}\right| <2$. In AM
this leads to a single conduction band $-2\leq E\leq 2$. In
KPM there exist multiple bands separated by band-gaps.
Allowed values of the energy variable belong to intervals 
\begin{equation}
k_{n}^{b}<k<\pi n,\text{ }n=1,2,3\ldots  \label{bands}
\end{equation}
where $k_{n}^{b}$ obeys the equation 
\begin{equation}
\tan \left( \frac{k_{n}^{b}a}{2}\right) =\frac{V_{0}}{2k_{n}^{b}},\text{ \ \ \ \ \ }n\text{ odd}
\label{boundodd1}
\end{equation}
\begin{equation}
\tan \left( \frac{k_{n}^{b}a}{2}\right) =-\frac{2k_{n}^{b}}{V_{0}}.\text{\ \ \ \ \ }n\text{ even}
\label{boundeven}
\end{equation}
The high energy boundaries of each band correspond to so called
resonances\cite{LGP} because disorder does not effect transport at these
particular energies. This fact can easily be seen from Eq. (\ref{width}),
where $\Gamma _{U}$ for KPM becomes zero for all $k=\pi n$. The
presence of these resonances is a specific property of the model under
consideration caused by the strict periodicity in the positions of site
potentials. Similar resonances are also present in other models such as the
dimer model\cite{dimer} or models of random superlattices.\cite{Tamura,Dima2}
The resonances dissappear once one destroys the exact periodicity in the
positions of $\delta $-functions or allows for random variations in the
width of superlattice's layers.

The main objects of our study are the finite size LE, $\tilde{\gamma}(L)$, and its variance
$\sigma^2$.  $\tilde{\gamma}(L)$ can be defined for the model under consideration as\cite{LGP} 
\begin{equation}
\tilde{\gamma}(L)=\frac{1}{L}\ln r_{N},  \label{LE}
\end{equation}
where $N=L/a$ is the total number of sites in the system and $r_{N}$ is the
envelope of the wave function 
\begin{equation}
r_{N}=\left( \psi _{N}^{2}+\psi _{N-1}^{2}\right) ^{1/2}.  \label{rN}
\end{equation}
As we discussed in the Introduction, $\tilde{\gamma}(L)$, 
takes a non-random value, $\gamma $, in the limit $L\rightarrow \infty $.
This limiting value can also be considered as an average of $\tilde{\gamma}%
(L) $ over different realizations of the system.\cite{Oseledets,LGP} For
large but finite $L$, $\tilde{\gamma}(L)$ exhibits finite size fluctuations
whose distribution function asymptotically approaches the Gaussian form with the
variance, $\sigma ^{2}$, decreasing as $1/L$.\cite{LGP,Furstenberg,Mello}

The average LE, $\gamma $, in the considered model was first
calculated in Ref.\onlinecite{Ishii}. It turns out that the
method developed in that paper (see also Ref.\onlinecite{LGP}) can be as well used
for exact calculation of $\sigma ^{2}$. The method is based upon the representation of LE in
terms of the phase variable $z_{n}$, defined as $z_{n}=\psi _{n}/\psi _{n-1}$,
which obeys the following equation of motion 
\begin{equation}
z_{n}+z_{n-1}^{-1}=U_{n}.  \label{z}
\end{equation}
Finite size LE can be expressed in terms of $z_{n}$: 
\begin{equation}
\tilde{\gamma}(L)=\frac{1}{L}\sum_{n=1}^{N}\ln \left| z_{n}\right| +\frac{1}{%
2L}\ln \left( r_{0}^{2}\frac{1+z_{N+1}^{2}}{z_{0}^{2}}\right) ,
\label{LEFinite}
\end{equation}
If $z_{n}$ is a stationary
random function of $n$, that is, a distribution of $z_{n}$ is independent of 
$n$, the first term in Eq. (\ref{LEFinite}) is of the order of unity while the
second term is of the order of ($1/L)$ and disappears in the limit $%
L\rightarrow \infty $. The expression for LE, therefore, takes the following
form 
\begin{equation}
\gamma =\lim_{L\rightarrow \infty }\frac{1}{L}\sum_{n=1}^{N}\ln \left|
z_{n}\right| =\left\langle \ln \left| z_{n}\right| \right\rangle,  \label{LEz}
\end{equation}
where the average on the right-hand side is taken over the stationary
distribution of $z$.

The asymptotic expression for variance of LE can be obtained from Eq. (\ref
{LEFinite}):
\begin{equation}
\sigma ^{2}=\frac{1}{L^{2}}\sum_{m,n=1}^{N}\left[ \left\langle \ln z_{m}\ln
z_{n}\right\rangle -\left\langle \ln z_{m}\right\rangle \left\langle \ln
z_{n}\right\rangle \right]  \label{variance}
\end{equation}
and is valid as long as the system's size $L$ is much greater than the correlation
radius of $z_{n}$, which we assume to be finite.

\section{Variance of the Lyapunov exponent in the Lloyd model}

\subsection{Two-point distribution of the phases $z_{n}$}

Calculation of the variance from Eq. (\ref{variance}) requires knowledge of
the two-point distribution function, $P_{2}(z_{n},z_{m})$, of the phases $z$. Our
calculations of this function are based upon representation of a joint
distribution of multiple random variables as the product of marginal and
conditional distributions: 
\begin{equation}
P_{2}(z_{n},z_{n+k})=P_{1}(z_{n})P(z_{n}|z_{n+k}),  \label{P2}
\end{equation}
where $P_{1}(z_{n})$ is a stationary probability distribution of $z_{n}$,
and $P(z_{n}|z_{n+k})$ denotes a conditional probability distribution of $%
z_{n+k}$ provided that $z_{n}$ is fixed. With the help of Eq. (\ref{z}) the
latter probability can be written as 
\begin{eqnarray}
P(z_{n}|z_{n+k})&=& \int \delta \left( z_{n+k}+z_{n+k-1}^{-1}-U_{n+k-1}\right)
\nonumber \\
&\times&P(z_{n}|z_{n+k-1},U_{n+k-1})dU_{n+k-1}dz_{n+k-1}, \nonumber
\end{eqnarray}
where $P(z_{n}|z_{n+k-1},U_{n+k-1})$ is a joint probability of $z_{n+k-1}$
and $U_{n+k-1}$. It follows from the structure of Eq. (\ref{z}) that 
$z_{n}$ depends only upon values of the random parameter $U_m$ at preceding sites $m<n$,
and thus is independent of $U_{n}$. The joint probability $%
P(z_{n}|z_{n+k-1},U_{n+k-1})$, therefore, can be factorized and integration
over $U_{n+k-1}$ can be carried out. The result is the following recurrent
relation between $P(z_{n}|z_{n+k})$ and $P(z_{n}|z_{n+k-1})$: 
\begin{eqnarray}
P(z_{n}|z_{n+k})&=& 
\int P\left( z_{n}|z_{n+k-1}\right) \label{Prec} \\
&\times& P_{C}\left(
z_{n+k}+z_{n+k-1}^{-1}\right) dz_{n+k-1},  \nonumber
\end{eqnarray}
where $P_{C}$ is the Cauchy distribution introduced in Eq. (\ref{P}). The
advantage of the Cauchy distribution is that recurrence (\ref{Prec}) can be
solved exactly. The conditional probability obtained has again the form of
the Cauchy distribution, which can be conveniently presented in the form 
\begin{equation}
P(z_{n}|z_{n+k})=\frac{{\rm Im}\xi _{k}}{\pi }\frac{1}{\left( z_{n+k}-\xi
_{k}\right) \left( z_{n+k}-\xi _{k}^{\ast }\right) },  \label{Pcond}
\end{equation}
where the asterisk denotes complex conjugation and parameters $\xi _{k}$ obey the following equation 
\begin{equation}
\xi _{k}+\xi _{k-1}^{-1}=U_{0}+i\Gamma .  \label{ksi}
\end{equation}
Eq. (\ref{Pcond}) for $P(z_{n}|z_{n+k})$ and Eq. (\ref{ksi})
for $\xi _{k}$ have exactly the same form as those obtained in Ref. 
\onlinecite{Ishii} for the one-point distribution $P_{1}(z_{n})$. However, in the
case of the one-point distribution one looks for a stationary solution of
Eq. (\ref{ksi}), while the conditional distribution $P(z_{n}|z_{n+k})$
requires that Eq. (\ref{ksi}) is solved with the initial condition 
\begin{equation}
\xi _{0}=z_{n}.  \label{InitCond}
\end{equation}
This solution can be presented as 
\begin{equation}
\xi _{k}=\frac{\delta ^{k}-\delta ^{-k}-z_{n}\left( \delta ^{k+1}-\delta
^{-k-1}\right) }{\delta ^{k-1}-\delta ^{-k+1}-z_{n}\left( \delta ^{k}-\delta
^{-k}\right) },  \label{ksisol}
\end{equation}
where $\delta $ is the $k$-independent solution of Eq. (\ref{ksi}), which obeys the stationary version of Eq. (\ref{ksi})  
\begin{equation}
\delta +\delta ^{-1}=U_{0}+i\Gamma .  \label{ksistateq}
\end{equation}
Real and imaginary parts of $\delta $ determine the center and the width of
the one-point distribution $P_{1}(z_{n})$%
\begin{equation}
P_{1}(z_{n})=\frac{{\rm Im}\delta }{\pi }\frac{1}{\left( z_{n}-\delta
\right) \left( z_{n}-\delta ^{\ast }\right) },  \label{P1}
\end{equation}
Averaging Eq. (\ref{LEz}) with the probability distribution (\ref{P1}),
one obtains the average LE, $\gamma $,\cite{LGP,Ishii} 
\begin{equation}
\gamma =\ln \left| \delta \right| /a  \label{LEres}
\end{equation}
Eqs. (\ref{Pcond}), (\ref{ksisol}), and (\ref{P1}) determine the two-point
probability distribution $P_{2}(z_{n},z_{n+k})$ defined by Eq. (\ref{P2}).

\subsection{Variance of the Lyapunov Exponent. General expression}

Eq. (\ref{variance}) for the variance of LE, $\sigma ^{2}$, can be presented
in the following form: 
\begin{eqnarray}
\sigma ^{2}&=&\frac{2}{L^{2}}\sum_{n=0}^{N-1}\sum_{k=1}^{N-n}\left\langle \ln
\left| z_{n}\right| \ln \left| z_{n+k}\right| \right\rangle \\ \label{variancereduced}
&+&\frac{1}{L}%
\left\langle \ln ^{2}\left| z_{n}\right| \right\rangle -\gamma ^{2}.
\nonumber
\end{eqnarray}
The correlation function $D(k)=\left\langle \ln \left| z_{n}\right| \ln \left|
z_{n+k}\right| \right\rangle $ is independent of the initial site $n$. With
the use of the probability distribution $P_{2}(z_{n},z_{n+k})$ found in the
previous subsection, it can be presented as 
\begin{equation}
D(k)=\frac{{\rm Im}\delta}{\pi }%
\displaystyle\int %
\limits_{-\infty }^{\infty }\frac{\ln \left| z\right| \ln \left| \xi
_{k}(z)\right| }{\left( z-\delta \right) \left( z-\delta ^{\ast }\right) }dz,
\label{D(k)}
\end{equation}
where $\xi _{k}(z)$ and $\delta $ are defined by
Eqs. (\ref{ksisol}) and (\ref{ksistateq}), respectively. Interchanging the
order of integration and summation in Eq. (\ref{variancereduced}) one can
find for the variance 
\begin{eqnarray}
\sigma ^{2} &=&\frac{2{\rm Im}\delta}{\pi aL}%
\displaystyle\int %
\limits_{-\infty }^{\infty }\frac{\ln \left| z\right| \ln \left| z\delta
-1\right| }{\left( z-\delta \right) \left( z-\delta ^{\ast }\right) }dz
\label{varint} \\
&-&\frac{1}{aL}\varphi (\pi -\varphi )-\frac{2}{L}\gamma \ln \left( \delta
^{2}-1\right) +O(1/L^{2}),  \nonumber
\end{eqnarray}
where $\varphi $ is the phase of $\delta $: 
\[\delta =p\exp (i\varphi ),\]
reduced to the interval $\left[ 0,\pi \right] $ and $p$ denotes the
absolute value of $\delta $: $\delta =$ $\left| p\right| $. The remaining
integral in Eq. (\ref{varint}) can be further simplified with the use of an
appropriate contour in the complex $z$-plane. One, finally, arrives at the
following expression for the variance in the case $\varphi <\pi /2$, which
corresponds to $U_{0}>0$. 
\begin{eqnarray}
&&\sigma ^{2} =\frac{1}{L}\left\{ -\gamma \ln \left[  
2{\displaystyle{\frac{\cosh(2\gamma a ) -\cos(2\varphi)}{\sinh^2(\gamma a )}}}\right]
\right.  \label{varfinal} \\
&& + \left.\frac{1}{a}%
\displaystyle\int %
\limits_{\varphi }^{\pi }dx\tan^{-1} \left[ \frac{\sinh(2\gamma a )\sin\varphi}{\cosh(2\gamma a )\cos \varphi
-\cos x} \right] \right\} +O\left( 1/L^{2}\right) ,  \nonumber
\end{eqnarray}
Since our model is symmetric with respect to the transformation $\varphi
\rightarrow \pi -\varphi $, the variance for $\varphi >\pi /2$ can be easily
evaluated.

\section{A new criterion for the single parameter scaling}

The necessary (but not sufficient) condition for SPS to hold is that the localization length be greater than all microscopic scales in the system. Therefore, we should consider the general result, Eq. (\ref{varfinal}), in the limit of large localization length, $\gamma a\ll 1$. 
Our first goal is to develop an asymptotic form of the integral in Eq. (\ref
{varfinal}) in this limit.
This is not a trivial task since the
integral has a singularity at $\gamma =0$. The first term in Eq. (\ref{varfinal}) is also singular at
this point, and one
would anticipate two singularities to cancel out. The latter singularity
has a logarithmic nature, $\gamma \ln \gamma $, and we need, therefore, to
extract the similar logarithmic singularity from the integral in Eq. (\ref
{varfinal}). To this end, we first evaluate the integral in Eq. (\ref
{varfinal}) by parts and present it in the following form: 
\begin{eqnarray}
\Phi (\gamma,\varphi ) &=&\pi \tan^{-1} \left( 
{\displaystyle{\beta  \over 1+\zeta  }} \right)
-\varphi \tan^{-1}\left( 
{\displaystyle{\beta \cos \varphi \over \zeta -\cos \varphi }} \right)
  \label{Phi} \\
&+&\beta 
\displaystyle\int %
\limits_{-1-\zeta  }^{-(\zeta -\cos \varphi) }dx\frac{\arccos %
\left( x+\zeta  \right) }{x^{2}+\beta ^{2}},  \nonumber
\end{eqnarray}
where
parameters $\beta $ and $\zeta $ are defined as
\begin{equation}
\beta =\sinh(2\gamma a) \sin \varphi ,  \label{beta0}
\end{equation}
\begin{equation}
\zeta =\cosh(2\gamma a)\cos \varphi.  \label{gamma}
\end{equation}
Since we are interested in small values of $\gamma a$, we can  expand $\cos^{-1} (x+\zeta) $ from Eq. (\ref
{Phi}) in the Taylor's series in $x$: 
\begin{equation}
\cos^{-1} ( x+\zeta) =\cos^{-1} (\zeta)+ \sum_{n=1}^{\infty }a_{n}x^{n}
\label{arccos}
\end{equation}
and carry out the term-wise integration. 
As the result we obtain 
\begin{eqnarray}
\Phi (\gamma,\varphi ) &= &\cos^{-1} ( \zeta  ) 
\left\{ 
\tan^{-1} 
\left( 
\frac{1+\zeta }{\beta } 
\right) 
\right. \\ \label{Phi1}
&-& \left. 
\tan^{-1} 
\left( 
\frac{ \zeta
-\cos \varphi) }{\beta }
\right) 
\right\}  \nonumber \\
&-&\frac{1}{2}\frac{\beta }{\sqrt{1-\zeta ^2}}\ln 
\left[ 
\frac{( \zeta -\cos \varphi)^2 +\beta^2}{( 1+\zeta )^2 +\beta^2} 
\right]  \nonumber \\
&+&\beta 
\left[ 
F(-(\zeta -\cos \varphi))-F(-1-\zeta )
\right] ,  \nonumber
\end{eqnarray}
where $F(x)$ is defined by the following power series 
\[
F(x)=\sum_{n=2}^{\infty }a_{n}\frac{x^{n-1}}{n-1}. 
\]
Comparing this expression with the original series (\ref{arccos}) one can
obtain for $F(x)$ the following integral representation 
\begin{equation}
F(x)=\frac{1}{x}\cos^{-1} ( \zeta ) +\frac{\ln \left|
x\right| }{\sqrt{1-\zeta ^{2}}}+\int \frac{dx}{x^{2}}\cos^{-1} (
\zeta  +x) .  \label{F(x)}
\end{equation}
The remaining integral in Eq. (\ref{F(x)}) can be calculated exactly. As a result we have
\begin{eqnarray}
F(x)&=&\frac{1}{x}\left[\cos^{-1} ( \zeta ) -\cos^{-1} ( \zeta+x )\right] \label{F(x)final} \\
&+&\frac{1}{\sqrt{1-\zeta ^{2}}}
\ln \left|1 - \zeta^2 (\zeta+x) \right. \nonumber \\
&+& \left. \sqrt{(1-\zeta^2)[1-(\zeta+x)^2]} \right|  .  \nonumber
\end{eqnarray}
When one combines Eq. (\ref{varfinal}) with Eqs. (\ref{Phi1}) and (\ref{F(x)final}),
the logarithmic singularity in $\Phi (\gamma,\varphi )$ nicely cancels out the
singularity in the first term of Eq. (\ref{varfinal}). The expression for the
variance $\sigma^2 $ emerging in the leading in $\gamma $ order takes the SPS
form : 
\begin{equation}
\sigma ^{2}\simeq 2\gamma/ L.  \label{SPS1}
\end{equation}
This is, to the best of our knowledge,  the first truly microscopic derivation of SPS with no {\em ad hoc}
hypotheses. The main reward for this is the exact criterion for SPS, which
follows from the conditions under which we have arrived at Eq. (\ref{SPS1}). First of all we assumed that
\begin{equation}
\frac{\beta }{\sqrt{1-\zeta ^{2}\cos ^{2}\varphi }}\ll 1. \label{ineq}
\end{equation}
Since
\begin{equation}
\beta \simeq 2\gamma a\sin \varphi,  \ \ \ \  \zeta \simeq 1+2(\gamma a)^{2}, \nonumber
\end{equation}
the inequality (\ref{ineq}) can be recast in the form 
\begin{equation}
\gamma a\ll \sin \varphi .
\end{equation}
Another condition, which we have to impose in order to
obtain Eq. (\ref{SPS1}) is 
\begin{equation}
(\zeta -1)\cos \varphi /\beta \simeq a\gamma
/\tan \varphi \ll 1. 
\end{equation}
Since $\sin \varphi \leq \tan \varphi$ the first of
the two inequalities is more restrictive, and the final condition for SPS
takes the form 
\begin{equation}
\kappa =l_{loc}/l_{s}\gg 1,  \label{criterion}
\end{equation}
where $l_{loc}=\gamma ^{-1}$ is the localization length, and a new length $l_{s}$
is defined as 
\begin{equation}
l_{s}=a/\sin \varphi.  \label{sparse}
\end{equation}

Evaluation of the integral in Eq. (\ref{varfinal}) in the limit $\kappa \ll 1$
can be performed by means of a simple expansion of the integrand in power
series in $\beta $ and retaining only the linear in $\beta $ term. The
resulting expression for $\sigma $ can be presented as 
\begin{equation}
\sigma ^{2}=
{\displaystyle{1 \over l_{s}L}}%
\left( \pi -%
{\displaystyle{2l_{loc} \over l_{s}}}%
\right).  \label{NSPS}
\end{equation}
Eq. (\ref{NSPS}) shows that in the regime considered here, $\sigma^2$ is
determined by the new length $l_{s}$ rather than by $l_{loc}$. It is important
to emphasize that in this limit both lengths $l_{s}$ and $l_{loc}$ can far 
exceed $a$, and Eq. (\ref{NSPS}), therefore, describes the violation of SPS
while the system remains within a meaningful scaling regime.

It should be noted, however, that Eq. (\ref{SPS1}) differs from Eq. (\ref{SPS}%
) by the factor of $2$. This discrepancy is due to the peculiar nature of
the Cauchy distribution, whose moments, starting from the second one, diverge.
Because of this, none of the standard approaches, used to derive Eq. (\ref{SPS})
within the random phase hypothesis, can be applied to the Lloyd model. In order
to illustrate this point, let us consider, for example, an expression for $%
\sigma ^{2}$ obtained in Ref.\onlinecite{Stone} for AM: 
\begin{eqnarray}
\sigma ^{2}&=&\frac{1}{N}
\left\{ 
Var
\left[ 
\ln 
\left( 
1+\frac{\epsilon ^{2}}{%
4-E^{2}}
\right) 
\right] \right.\label{VarStone} \\
&+& \left.\left\langle \left( \ln \left[ 1+2\frac{\epsilon }{%
\sqrt{4-E^{2}}}\cos \nu +\frac{\epsilon ^{2}}{4-E^{2}}\right] \right)
^{2}\right\rangle \right\}.   \nonumber
\end{eqnarray}
Brackets $\left\langle {}\right\rangle $ designate here averaging over the
random site energy $\epsilon $ and the phase $\nu $, which is assumed to be
statistically independent of $\epsilon $, and distributed uniformly. $Var$
denotes the variance of the respective quantity. The standard weak
disorder expansion used in Ref.\onlinecite{Stone} implies the expansion of this
expression in powers of the random variable $\epsilon $ with consecutive
averaging. The first term in Eq. (\ref{VarStone}) then becomes of the order
of $\left( \left\langle \epsilon ^{2}\right\rangle /\left[ 4-E^{2}\right]
\right) ^{2}$ and is neglected, while the second term after averaging over
the phase yields Eq. (\ref{SPS}). In the case of the Cauchy distribution for 
$\epsilon$, this approach cannot be applied because $\left\langle \epsilon
^{2}\right\rangle $ does not exit. In order to pass to the weak-scattering
limit one has to average over $\epsilon $ first, and only after that carry
out expansion over the parameter $\Gamma $ of the Cauchy distribution. Both terms in
Eq. (\ref{VarStone}) then become of the same order of magnitude $\Gamma /%
\sqrt{4-E^{2}}$, and though the general proportionality between $\sigma ^{2}$
and LE is preserved, the numerical factor not equal to that
of Eq. (\ref{SPS}) or of Eq. (\ref{SPS1}). This result implies that the phase
randomization hypothesis is not valid at all for the Lloyd model. What is more
important, however, is the fact that although the phase
randomization hypothesis fails, SPS still survives.

For $\kappa \ll 1$, one can provide a clear physical interpretation
for the length $l_{s}$. According to Thouless, \cite{Thouless} the phase $%
\varphi (E)$ is proportional to the integrated density of states $%
G(E)=\varphi (E)/\pi$, and $\kappa \ll 1$ corresponds to either $%
G(E)\ll 1$ or $1-G(E)\ll 1$. The length $l_{s}$ then can be expressed
in terms of the number of states in the energy intervals  between $E$ and the
closest  boundary of the spectrum $l_{s}=1/(\pi G(E))$ [or $%
l_{s}=a/(\pi -a\pi G)$]. For AM these boundaries lie at $\pm \infty $, and
for KPM they are the resonance boundaries of the bands, where $\varphi (E)=\pi n$ with $n$
an integer. The states in these regions arise due to rare realizations of the
disorder, and can be associated with spatially localized and well-separated
structural defects. The length $l_{s}$ then can be interpreted as an average
distance between such defects. In view of this interpretation of $l_{s}$,
the physical meaning of the transition between two types of scaling regimes
also becomes clear. Condition (\ref{criterion}) means that the localization 
length at the energy $E$ exceeds the spatial separation between neighboring 
localization centers from the relevant part of the tail (between $E$ and the 
nearest boundary of the spectrum).  Under this condition the localized states
overlap and SPS is valid.

To complete our
discussion of the new scaling parameter $l_{s}$, let us compare it with
the phase randomization length, $l_{ph}$ numerically studied  by Stone, et al., Ref.\onlinecite
{Stone}. Assuming that $E$ is inside the conduction band and far from the band boundaries,
we can approximate $l_{s}$  as 
\begin{equation}
l_{s}^{-1}\simeq \frac{1}{2a}{\sqrt{4-U_{0}^{2}}}+O(\Gamma _{U}^{2}).
\label{phaseCB}
\end{equation}
According to Eq. (\ref{phaseCB}), $l_{s}$ decreases toward the center of the
band $U_{0}=0$, where it reaches its minimum value equal to $a$. At the same
time, the phase randomization length was found in Ref.\onlinecite{Stone} to increase
toward the center of the band $E=0$, where it seemed to diverge. For $E=0$,
the phase distribution was found to be nonuniform even for very long chains. The
absence of the phase randomization in the center of the band was also found
analytically in Ref.\onlinecite{Dmitriev}. This comparison proves that $l_{s}$ is
an independent new parameter responsible for the statistics of LE. Both
numerical results of Ref.\onlinecite{Stone} and analytical calculations of Ref.
\onlinecite{Dmitriev} show that a nonuniform distribution of phases can be
consistent with SPS, providing an additional argument against the condition
for SPS based upon phase randomization. At the same time, our criterion,
Eq. (\ref{criterion}), correctly predicts validity of SPS in the band center
as long as the localization length remains macroscopic.

\subsection*{$l_s$ and the SPS criterion in generic models}

The peculiarities of the Lloyd model may cast doubts on the robustness of the new scale $%
l_{s}$ and the criterion Eq. (\ref{criterion}). In order to show that this criterion 
is applicable beyond the Lloyd model, we carry out additional numerical simulations 
of the model studied previously in Ref.\onlinecite
{Dima2}. That model is of the Kronig-Penney kind, but unlike the model considered
in the present paper, its potential is formed by rectangular barriers. The
width of the barriers is assumed to be random with a uniform distribution
over a given interval. Both the potential and the statistics of the model
used in numerical simulations is considerably different from the Lloyd
model, e.g., all moments of the distribution function exist. 

It is instructive to rewrite expressions for $\sigma ^{2}$ in terms of a new
dimensionless variable $\tau =\sigma ^{2}Ll_{loc}/2$ as a function of $%
\kappa $. In terms of these variables both asymptotics of the variance, $\sigma^2$ given
by Eqs. (\ref{SPS1}) and (\ref{NSPS}), can be presented in a form which
contains no free parameters: 
\begin{equation}
\tau =\left\{ 
\matrix{
\kappa \left( 
{\displaystyle{\pi  \over 2}}-\kappa \right) & \kappa \ll 1 \cr 
1&\kappa \gg 1}
\right.   \label{universal}
\end{equation}
Although we do not expect the concrete form of the function $\tau (\kappa )$ to be
universal, we do believe that  the new crossover length $l_{s}$ retains its physical meaning in the general case,  and that the crossover point is also universally determined by $\kappa\sim 1$. 

In order to generalize  the  crossover length $l_s$ for other models, we use the interpretation of the
phase $\varphi $ in terms of the integral density of states normalized in such a way that the phase would vary between $0$ and $\pi$ for any given band. The generalization is quite straightforward for models with a single band spectrum if in the absence of disorder the band has a finite width. The total number of states in such models is finite, and it can be used to normalize the phase. If the initial band of the system is infinitely broad, e.g., for the Schr\"{o}dinger equation with a random white-noise potential,  one has to introduce a cut-off frequency for the spectrum in order to normalize the phase. The
crossover parameter $\kappa $ for the later model can be obtained from an expression
relating LE and the integral density of states  found in Ref.\onlinecite{LGP}:
\begin{equation}
\kappa ^{-1}\left( E\right) =\frac{\sqrt{\pi }}{2}\int_{0}^{\infty }\sqrt{x}%
\exp \left( -\frac{x^{2}}{12}-\frac{Ex}{D^{2/3}}\right) dx, \label{gauss}
\end{equation}
where $D$ determines the strength of the $\delta $-correlated potential.  The asymptote of $\kappa $
for large negative $E$ is 
\[
\kappa \approx \exp \left( -\frac{4\left| E\right| ^{3/2}}{3D}\right) \ll 1,
\]
and we conclude that these states do not obey SPS. Transition to SPS
behavior again occurs at the initial band boundary $E=0$, where $\kappa
\approx 1.1$ and does not depend upon parameters of the model. 

For systems with multiple (in the absence of disorder) bands separated by band-gaps, one has to consider separately two different situations. If disorder-modified bands still do not overlap and a genuine gap between the bands persist, the situation is equivalent to the single-band case. The phase can be defined for each separate band and normalized by the number of states in the band. 
The results of the numerical simulations of this particular situation are shown in Fig. 1 along with $%
\tau (\kappa )$ obtained from our analytical Eq. (\ref{varfinal}). $l_s$ was calculated numerically  with the phase defined using the integral density of states normalized in such a way that the phase changes from zero to $\pi$ when energy sweeps over a band from one fluctuation boundary to the other. One can
see from this figure that the crossover between different asymptotes  for both numerical and analytical calculations occurs in the same region. This proves the universal significance of our criterion (\ref{criterion}) for
SPS and justifies the suggested generalization of the crossover length $l_{s}$. 
\begin{figure}
\centering
\vspace{-0.15in}
\epsfxsize=3in \epsfbox{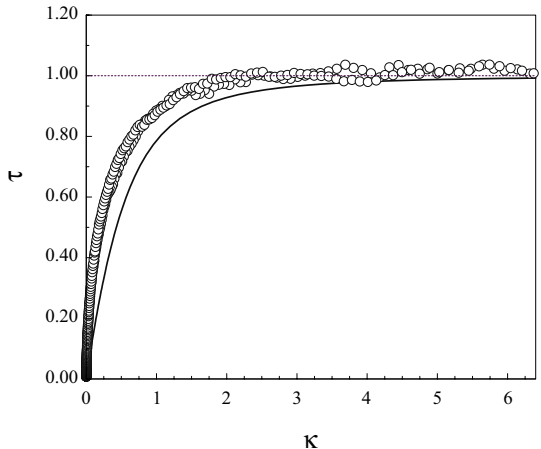}
\vspace{-0in}
\caption{The function $\tau(\kappa)$ obtained from the analytical solution of the Lloyd model Eq. (\ref{varfinal}) (solid line) and the numerical simulations of KPM with rectangular barriers of  random widths (open circles). Note that the SPS equations for these models differ by a factor of 2 [Eqs. (\ref{SPS}) and (\ref{SPS1})]; therefore we rescaled numerical data by this factor.
}
\end{figure}
A new situation arises, however, when fluctuation states from adjacent  bands overlap, and the spectrum does not have boundaries in former band-gaps. On one hand, it is clear that if the number of states in the former gap is small, SPS should not be expected on the basis of the general qualitative interpretation of the criterion (\ref{criterion}). On the other hand, since there are no exact boundaries of the spectrum inside former gaps, one cannot define a phase suitable for determination of $l_s$. Though this situation requires special consideration, we can offer a conjecture that can be used to meaningfully define $l_s$ in this case. Consider one of the original bands between two adjacent band-gaps. In the presence of disorder there appear tails of the density of states within the gaps. When disorder is small, one can always distinguish between gap states originating from different bands (except for a small region where the tails from different bands overlap). Let $g_n(E)$ be the differential density of states related to the $n$-th band. Then integral $N_{tot} = \int_{-\infty}^{\infty} g_n(E)dE$ gives the total number of states originating from the band. One can define a phase 
\[
\varphi_n(E)=\pi\int_{-\infty}^Eg_n(E)dE/N_{tot},
\]
with $E$ obeying the following inequality $E_{min}^{n-1}<E<E_{min}^{n}$, where $E_{min}^n$ corresponds to the minimum of the {\it actual} differential density of states within the gap between the $n$-th and $(n+1)$-th bands.  The phase defined according to this procedure does not assume unphysical values of $\pi$ at the points where there are no spectrum boundaries, and we suggest that the parameter $l_s$ defined through this phase according to
\begin{equation}
l_s^{-1} = \sin\left[\frac{\pi}{N_{tot}}\int_{-\infty}^Eg_n(E)dE\right], \hspace{10pt} E_{min}^{n-1}<E<E_{min}^{n},
\end{equation} 
can be used in order to formulate the criterion Eq. (\ref{criterion}). The suggested definition of $l_s$ can be practically used for analytical estimates of the transition between different statistics, using, for example, a tight-binding approach to a multi-band problem, where interactions only between adjacent bands are taken into account. However, more detailed discussion of this issue requires a separate paper. 

\section{Properties of the transition region between SPS and non-SPS states}

In this section we discuss properties of the transition region
between SPS and non-SPS states.  In spite of the mentioned peculiarities
of the Lloyd model, our calculations provide a sound qualitative explanation for
numerical results of Ref. \onlinecite{Dima2}, confirming once again that we correctly describe
the qualitative nature of the transition between SPS and non-SPS
statistics. In the model considered, the phase, $\varphi$, and LE, $\gamma $, can
be conveniently presented in the following form\cite{LGP} 
\begin{equation}
\sin \varphi =\frac{\sqrt{s}}{\sqrt{U_{0}^{2}+s}},  \label{phi}
\end{equation}
\begin{equation}
\gamma \simeq \Gamma _{U}/\sqrt{s},  \label{gamma0}
\end{equation}
where $s$ is given as
\begin{equation}
s=\frac{1}{2}\left( 4+\Gamma _{U}^{2}-U_{0}^{2}\right) +\frac{1}{2}\sqrt{%
\left( 4+\Gamma _{U}^{2}-U_{0}^{2}\right) ^{2}+4U_{0}^{2}\Gamma _{U}^{2}}
\label{s}
\end{equation}
and we assume again that $\gamma \ll 1$. The relation between $\gamma $ and 
$\varphi $ is determined by the parameter $s$, which in its turn depends upon
the energy $E$. Let us recall that the energy enters into our equations through
parameters $U_{0}$ and $\Gamma _{U}$ defined in Eqs. (\ref{UAV}) and  (\ref
{width}). For energies within conduction bands, LE is of the order of $\Gamma
_{U}$ though $l_{s}$ is of the order of one [see Eq. (\ref{phaseCB})]. Thus,
SPS holds as long as disorder is small, $\Gamma _{U}\ll 1$, in accordance
with the previous results.\cite{Anderson,Abrikosov,Shapiro,Stone} From Eqs. 
(\ref{phi}), (\ref{gamma0}), and (\ref{s}) it is clear that the relation
between $l_{loc}$ and $l_{s}$ changes with the energy approaching an initial
spectral boundary. In the limit of small disorder $l_{loc}=l_{s}$ exactly at
the boundary. Therefore, one should expect the strongest violation of SPS
for states which arise due to disorder in the originally forbidden regions.
For AM this corresponds to energies $\left| E\right| >2$, and
for KPM these are energies from band-gaps of the original spectrum. For energies
liying far away from the boundaries one can obtain the following approximate
expressions for LE and the phase $\varphi $.

\begin{eqnarray}
\sin \varphi &\sim &\frac{\Gamma _{U}}{\sqrt{U_{0}^{2}-4}}\ll 1,
\label{phaseinside} \\
\gamma &\sim &\frac{\sqrt{U_{0}^{2}-4}}{U_{0}}.  \label{lambdainside}
\end{eqnarray}
It is evident that in this case $l_{loc}\ll l_{s}$, and the variance behaves
according to Eq. (\ref{NSPS}). The states disobeying SPS, however, are more
important for KPM than for AM. The
reason for this is that LE in AM becomes of the order of one not
very far from the boundary, moving the system out of any scaling regime. In
KPM $\gamma $ can remain small enough throughout entire
band-gaps for sufficiently high energies $k\gg V_{0}$, and the violation of
SPS in this case occurs when the localization length is still of a macroscopic
scale.

Eqs. (\ref{SPS1}) and (\ref{NSPS}) explain a non-monotonic behavior of $\sigma
(E)$ observed numerically in Ref.\onlinecite{Dima2}. When the energy moves towards a
band edge, LE grows and $\sigma $ grows along with $\gamma $. When,
however, $\gamma $ becomes equal to $l_{s}^{-1}$ the variance,  $%
\sigma^2 $ starts decreasing towards the value  $\sim \pi
l_{s}^{-1}/N.$ The maximum of $\sigma $, therefore, corresponds to the
energy where $\gamma \simeq l_{s}^{-1}$, i.e. the boundary of the original
spectrum.

We can now also estimate the width of the transition region between SPS and
non-SPS states, which was found in numerical simulations to be surprisingly
small. The transition between the two groups of states occurs when $\sigma
(E)$ passes through its maximum, and the width of the transition region is
related to the sharpness of the maximum. In view of the
preceding discussion, the latter is determined by the region of energies over which $%
\gamma (\Gamma _{U})$ changes its behavior. The extent of this region can be
estimated from the condition 
\begin{equation}
\left| 4-U_{0}^{2}\right| \sim 2U_{0}\Gamma _{U}
\end{equation}
In AM it leads to $\delta E\sim \Gamma $, in KPM one has 
\begin{equation}
\delta k\sim \frac{\Gamma }{ak_{n}\delta _{n}}\sim \frac{\Gamma }{V_{0}}%
\frac{\Delta _{n}}{\delta _{n}},  \label{sharpness}
\end{equation}
where $\Delta _{n}$ is the width of $n$-th band-gap, $k_{n}$ represents the non-resonant boundary of the $n$-th band, and
the parameter $\delta _{n}$ is defined as 
\begin{equation}
\delta _{n}=1+\frac{4 V_{0}}{ak_{n}^{2}}.  \label{deltan}
\end{equation}
 In both cases the width
of the transition region is determined by the degree of disorder in the
system, and is small when disorder is small. In the Kronig-Penney situation,
however, Eq. (\ref{sharpness}) indicates a special sharpness of the
transition in the case of high-energy bands, when the parameter $\Delta _{n}$ is
also small.

When disorder increases, AM and KPM behave differently. Monte Carlo results\cite{Dima2}
show that in periodic systems, an increase of disorder leads to
a restoration of SPS for almost the entire spectrum of the system. We are now 
able to explain this behavior and to provide an estimate for the critical
disorder. It is clear that the parameter $l_{s}^{-1}$ reaches a minimum at the energy
in the center of a band gap. This minimum value can be estimated from Eq. (\ref
{lambdainside}) as 
\begin{equation}
l_{s\min }^{-1}\sim \frac{\Gamma }{k_{n}}\sqrt{\delta _{n}},  \label{varmin}
\end{equation}
where $k_{n}$ represents the non-resonant boundary of the $n$-th band, and
the parameter $\delta _{n}$ is defined by Eq. (\ref{deltan}). 
At the same the energy where $l_{s}^{-1}$ is minimal, LE  assumes its
maximum value: 
\begin{equation}
\gamma _{\max }\sim \frac{V_{0}}{k_{n}\sqrt{\delta _{n}}}.
\end{equation}
$l_{s\min }^{-1}$ increases with disorder, while $\gamma
_{\max }$ does not change, and at $\Gamma \simeq V_{0}/\delta _{n}$ two length scales are of the same order,
$l_{s\min }\simeq $ $\gamma _{\max }^{-1}$. At this instant for the states
outside of the immediate vicinity of the center of the band-gap, 
$l_{s}\gg \gamma^{-1} $ and SPS is restored. Thus, we can identify $\Gamma
_{cr}^{n}=V_{0}/\delta _{n}$ as a critical disorder for the $n$-th band-gap. For
the states right in the center of the gaps, however, $l_{s}\sim \gamma^{-1}$
no matter what disorder is, and these states do not obey SPS. Therefore, the
complete restoration of SPS for the entire band-gap does not occur in this
model, but the width of the non-SPS region decreases with increase of
disorder, as one can see from Fig. 2.
\begin{figure}
\centering
\vspace{-0.15in}
\epsfxsize=2.8in \epsfbox{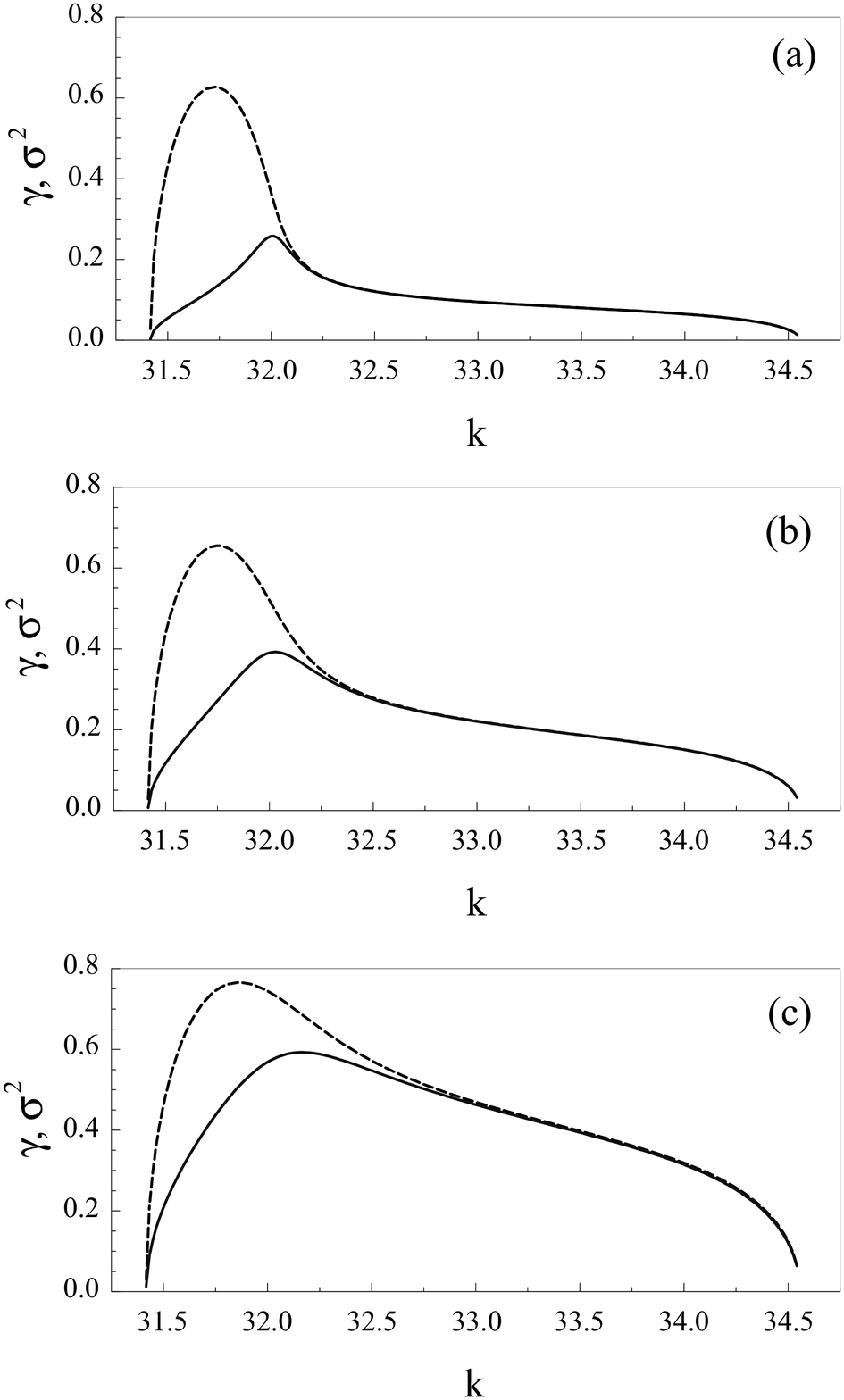}
\vspace{0.1in}
\caption{The variance of LE (solid line) and LE itself (dashed line) for
KPM are posted as functions of the energy parameter, $k$, for different degrees of disorder. The plot extends over the region $10\pi<k<11\pi$, so one gap and one conductance band are represented.  The presence of
the maximum in the vicinity of the band boundary ($k\sim 10.35\pi$) signals the violation of
SPS for states from a band-gap. The absence of a similar maximum at the
second boundary is due to its resonance nature.  The degrees of the disorder, $\Gamma$,  are 3 (a), 7 (b), and 15 (c).  It is clearly seen that
with the increase of the disorder, the number of states disobeying SPS
decreases.}
\end{figure}

In the case of a strong potential $V_{0}$ and in the high
energy,  $a^{-1} \ll V_{0}\ll k_{n\text{,}}$ one has $\delta _{n}\simeq 1$,
and the critical disorder is just equal to $V_{0}$. When a potential is weak,
$V_{0}a\ll 1$, there exists a medium-energy regime, when $\sqrt{V_{0}/a}\gg
k_{n}\gg V_{0}$. In this case 
\begin{eqnarray}
\gamma _{\max } &\sim &\sqrt{V_{0}a}\ll 1,  \nonumber \\
l_{s\min }^{-1} &\sim &\frac{\Gamma }{k_{n}^{2}}\sqrt{\frac{V_{0}}{a}},  \nonumber  \\
\Gamma _{cr}^{n} &\simeq &ak_{n}^{2}. \nonumber 
\end{eqnarray}
It is interesting to note that in this case $\Gamma _{cr}^{n}$ increases
with an increase of $k_{n}$, even though the widths of the gaps decrease.

The calculations presented above referred to the non-resonant band
boundaries. At the resonant points, $ka=\pi n$, both the localization length $%
l_{loc}$ and the crossover length $l_{s}$ diverges, while the variance
of LE vanishes. Although the resonances are not stable with respect to a
violation of the periodic arrangements of the $\delta $-potentials, they
occur in some other models as well. We already mentioned models with
correlated disorder\cite{dimer} and random superlattices.\cite{Tamura,Dima2}
The latter has an experimental significance with applications
to propagation of classical waves. Therefore, it is interesting to consider
the behavior of the critical parameter $\kappa $ in the vicinity of the
resonances. Although both $l_{loc}$ and $l_{s}$ 
diverge at the resonances, their ratio $\kappa $ remains finite and takes on
the following values: 
\begin{equation}
\kappa =\left\{ 
\matrix{
{\displaystyle{V_{0} \over \Gamma _{0}}}+\sqrt{1+{\displaystyle{V_{0} \over \Gamma _{0}}}},&ka<\pi n \cr 
-{\displaystyle{V_{0} \over \Gamma _{0}}}+\sqrt{1+{\displaystyle{V_{0} \over \Gamma _{0}}}}.&ka>\pi n}
\right.  \label{kappares}
\end{equation}
One can see from Eq. (\ref{kappares}) that $\kappa $ experiences a
discontinuity at resonance points: its value decreases by $2V_{0}/\Gamma
_{0}$ once a point is crossed. In the case of small disorder, when $%
V_{0}/\Gamma _{0}\gg 1$, this is a dramatic jump, such that $\kappa \gg 1$
at the band side of the resonant boundary and $\kappa \ll 1$ at the gap
side. It is obvious, therefore that scaling properties of the system also
change discontinuously at the resonance from SPS behavior at the band side to
the scaling with two parameters at the gap side.

\section{Conclusion}

In this paper we studied statistical properties of the Lyapunov exponent in the
one-dimensional Anderson model with the Cauchy distribution of site energies.
The model can also be interpreted as the Kronig-Penney model with periodically
positioned $\delta $-potentials with random strengths. The main objective of
the study was to find an exact solution for the thermodynamical limit
of the variance of LE and to establish an exact
criterion for the existence of single parameter scaling. It is important to
emphasize that in contrast with all previous calculations of the variance, we did
not use the phase randomization hypothesis. This allowed us to reject the
generally accepted assumption that it is the length over which the phase of
reflection and transmission coefficients becomes uniformly distributed that sets
the condition for the existence of SPS.

We found a new length scale, $l_{s}$, which is responsible for
the scaling properties of the conductivity in the system: SPS exists as long as
the localization length $l_{loc}$ exceeds $l_{s}$. The length $l_{s}$, however,
differs from the phase randomization length, and presents,
therefore, a new significant scaling parameter. The parameter $l_{s}$ is
microscopic for states close to the center of the original conduction
bands of the system and does not impose, therefore, any additional
restrictions for the existence of SPS excepting the regular requirement for
the localization length to be of a macroscopic dimension. However, for the
states at the edge of the bands, $l_{s}$ grows to a macroscopic size and
the condition $l_{loc}=l_{s}$ actually establishes a boundary between the states
with and without the SPS statistics. As soon as $l_{s}$ becomes much larger than 
all microscopic lengths, this scale becomes significant. In this limit it can be expressed in terms of the number of states $N_{def}(E)$ which arise at the tails of the initial bands due to rare
fluctuation configurations: $l_{s}^{-1}=N_{def}(E)$. It then can be given a
natural physical interpretation as an average distance between such defects.

The change of the scaling behavior occurs when the energy crosses over a
boundary of a former gap. In the case of regular boundaries, the change occurs
gradually with the critical parameter $\kappa$ being of the order of unity
right at the boundary. The Kronig-Penney version of our model, besides regular
boundaries has so called resonant boundaries, where both LE and 
$l_{s}^{-1}$ vanish. We found that at the resonance boundaries the parameter $%
\kappa $ undergoes a sudden jump from very large values $\kappa \gg 1$ at
the band side of the boundary to very small values $\kappa \ll 1$ at the gap
side. This means that the change of the scaling behavior at the resonant
energies also occurs discontinuously: the system obeys SPS when the boundary
is approached from the conduction band, and then demonstrates two
parameter scaling if the boundary is approached from the gap.

We carried out numerical simulations of
the Kronig-Penney-like model with a different configuration of the potential and
different statistics. The comparison between numerical and analytical results
clearly indicates that significance of the length scale $l_{s}$ defined in
terms of the integral density of states persists beyond the Lloyd model, and
that the new criterion for SPS established in the present paper has a
universal nature.

\section*{Acknowledgments}
We are indebted to A. Mirlin for a useful discussion. We also wish to thank S. Schwarz for reading and commenting on the manuscript.  Work at Seton Hall University was supported by NATO Linkage Grant N974573, work at Queens College was supported by a CUNY collaborative grant and PSC-CUNY research award, and work at Princeton University was supported by ARO under contract DAAG 55-98-1-0270.

\end{multicols}

\begin{references}
\bibitem{note1} It should be noted,
however, that often the factor of $1/2$ in the definition of LE (\ref{gamma1}) is omitted,
Eq. (\ref{SPS}) then contains an additional factor of 2 at its right-hand side.
\end{references}
\end{document}